\documentclass[doublecol]{epl2} 
\usepackage{graphicx}
\usepackage{amsmath}
\usepackage{amssymb}
\usepackage{epstopdf}
\newcommand{\be}{\begin{eqnarray}}
\newcommand{\ee}{\end{eqnarray}}
\newcommand{\bse}{\begin{subequations}}
\newcommand{\ese}{\end{subequations}}

\newcommand{\bit}{\begin{itemize}}
\newcommand{\eit}{\end{itemize}}
\newcommand{\ben}{\begin{enumerate}}
\newcommand{\een}{\end{enumerate}}

\newcommand{\bpm}{\begin{pmatrix}}
\newcommand{\epm}{\end{pmatrix}}

\newcommand{\mcal}{\mathcal}

\newcommand{\mrm}{\mathrm}

\newcommand{\bs}{\boldsymbol}

\newcommand{\p}{\partial}
\newcommand{\f}{\frac}
\newcommand{\diff}{\mrm{d}}

\newcommand{\J}{\mrm{J}}
\newcommand{\MJ}{\mrm{MJ}}

\newcommand{\ga}{\alpha}
\newcommand{\gb}{\beta}
\newcommand{\gc}{\gamma}

\newcommand{\eps}{\epsilon}

\newcommand{\gr}{\varrho}

\newcommand{\Gs}{\Sigma}

\newcommand{\kB}{k_\mrm{B}}

\newcommand{\HamEn}{\mcal{E}}

\newcommand{\Temp}{\mcal{T}}
\newcommand{\dt}{\diff t}

\title{Stationarity, ergodicity, and entropy in relativistic systems}
\shorttitle{Stationarity, ergodicity \& entropy in relativity}

\author{David Cubero\inst{1} and J\"orn Dunkel\inst{2}}
\shortauthor{D. Cubero and J. Dunkel}
\institute{
 \inst{1} Departamento de F\'{\i}sica Aplicada I, EUP, Universidad de Sevilla, Calle Virgen de \'Africa 7, 41011 Sevilla, Spain \\
 \inst{2} Rudolf Peierls Centre for Theoretical Physics, University of Oxford, 1 Keble Road, Oxford OX1 3NP, United Kingdom}
\abstract{
Recent molecular dynamics simulations show that a dilute relativistic gas equilibrates to a J\"uttner velocity distribution if ensemble velocities are measured simultaneously in the observer frame. The analysis of relativistic Brownian motion processes, on the other hand, implies that stationary one-particle distributions can differ depending on the underlying time-parameterizations. Using molecular dynamics simulations, we demonstrate how this  relativistic phenomenon can be understood within a deterministic model system. We show that, depending on the time-parameterization, one can distinguish different types of \lq soft\rq\space ergodicity on the level of the one-particle distributions. Our analysis  further  reveals a close connection between time parameters and entropy in special relativity. {A combination of different time-parameterizations can potentially be useful in simulations that combine molecular dynamics algorithms with randomized particle creation, annihilation, or decay processes.}
}
\begin{document}

\maketitle

%\paragraph{Introduction.--}
{\em Introduction.--}
Understanding the relation between ensemble and time averages poses one of the most fundamental problems  in statistical physics. Ergodicity -- the equivalence of the two averaging procedures --   is a commonly employed assumption in statistical mechanics~\cite{Becker}, albeit difficult to prove for realistic systems. During the past decades, the ergodicity hypothesis was  intensely examined for nonrelativistic classical~\cite{1982Pa,1990PeLa,2005BeBa,2007ReBa} and  quantum models~\cite{1984StHe,1992Bo,1998KaHe}. However, much less is known about its meaning  and validity in relativistic settings~\cite{1970JoLa_1}, when even more basic concepts like \lq stationarity\rq~may become ambigous as time becomes relative~\cite{1967Ga,2009DuHaWe,2009DuHa}.   
A clear conception of the interplay between time parameters and thermostatistical concepts, like entropy~\cite{1970JoLa_2,2007DuTaHa_2,2006Ka}, is crucial, e.g.,  if one wishes to generalize non-equilibrium fluctuation theorems to a relativistic framework~\cite{2007Fi,2008ClEtAl}. Given the rapidly increasing number of applications in high-energy physics~\cite{2008BrEtAl,2008KaEtAl} and astrophysics~\cite{2006WoMe,2004ItNo}, a firm conceptual foundation is desirable not only from a theoretical, but also from a  practical perspective. 
\par
{
Ideally, one would like to tackle relativistic many-particle problems within a quantum field theory framework, as this allows for the consistent treatment of particle creation, annihilation, or decay processes~\cite{WeinbergQFT1}. In recent years, substantial progress has been made towards a better understanding
of both equilibrium~\cite{2001MoRo,2007BeFe} and non-equilibrium processes~\cite{2000AaBoWe,2001AaBoWe,2002BlIa,2005Be}  in the context of relativistic field theories. However, while without doubt conceptually preferable, an exact quantum theoretical treatment is in many situations practically unfeasible and, for a considerable number of applications (e.g., sufficiently dilute gases or plasmas), not necessary.   With regard to computer simulations of relativistic systems, suitably adapted quasi-classical particle models~\cite{1989SoStGr} often provide a more efficient basis for quantitive numerical analysis.  In particular, given the rapid improvement of GPU programming tools~\cite{2009PrEtAl,2009GeEtAl,2009JaKo,2009ThEtAl} over the past two years, relativistic molecular dynamics (MD) simulations as well as other parallelizable  approaches, e.g., particle based Monte-Carlo (MC) algorithms~\cite{2009PeEtAl}, can be expected to play an increasingly important role in the future. Against this background, the present paper addresses selected thermostatistical properties of quasi-classical relativistic systems.
}
\par
More precisely, we intend to demonstrate that even a relatively simple, relativistic model system~\cite{2007CuEtAl} may provide insights into basic conceptual questions, such as: How are observer-time and  proper-time averages of single-particle  trajectories related to each other? How are the resulting time-averaged distributions linked to stationary distributions obtained from simultaneous ensemble measurements?
Is it possible to establish  a connection between time parameters and entropy? 
{As we shall see, the answers also provide further clarification of results that were recently obtained in the theory of relativistic Brownian motions~\cite{1997DeMaRi,2005Zy,2007Li,2008ChDe,2009DuHa,2009DuHaWe}. For example, one can show that changing the time-parameterization of a relativistic stochastic process (e.g., form coordinate-time to proper-time)   entails a modification of the corresponding stationary distribution~\cite{2009DuHaWe}. Below, we will discuss how this phenomenon can be understood on the basis of a simple deterministic model system. 
Furthermore, we are going to illustrate how thermodynamic state variables become modified when replacing coordinate-time through a proper-time parameterization. Generally, proper-time parameterizations provide a natural framework for including particle decay processes in relativistic simulations, whereas global coordinate-time parameterizations are better suited for quantifying the  many-particle dynamics and, in particular, the causal ordering of collision events~\cite{1989SoStGr}. Thus, as briefly outlined in the latter part of this paper, a combination of different time parameterizations may yield useful mixed  MD/MC-simulation schemes.
}

%\paragraph{Time-averaged single-particle distributions.--}
{\em Time-averaged single-particle distributions.--}
Let us start by considering the motion of a specific particle in an inertial frame $\Gs_0$. The velocity $\bs V:=\diff \bs X/\dt$ of the particle can be parameterized in terms of the $\Gs$-coordinate-time~$t$,  denoted by $\bs V(t)$, or, alternatively, by the particle's proper-time (units such that the speed of light $c=1$)
\be
\tau(t)=\int_0^t\,dt' \sqrt{1-\bs {V}(t')^2},
\ee
corresponding to a function $\hat{\bs V}(\tau)$ that satisfies $\bs V(t)=\hat{\bs V}(\tau(t))$.  We may then define the $t$-averaged velocity probability density function (PDF) of the particle by
\bse
\be
f_t(\bs  v)=
%\lim_{t_+\rightarrow\infty}
\f{1}{t} \int_0^{t} dt'\, \delta[{\bs v}-\bs V(t')],
\ee
and, similarly, the associated $\tau$-averaged PDF by
\be
\hat{f}_\tau({\bs  v})=
%\lim_{\tau\rightarrow\infty}
\frac{1}{\tau}
\int_0^{\tau} d\tau'\, \delta[{\bs v}-\hat{\bs V}(\tau')].
\label{eq:fproper}
\ee
\ese
We would like to understand how the two PDFs $f_t$ and $\hat{f}_\tau$ are related to each other as $t,\tau\to\infty$.  
To this end, we change the integration variable in Eq.~\eqref{eq:fproper} to the lab time,  yielding 
\be\label{e:relation}
\hat{f}_\tau({\bs  v})
=
\f{t}{\tau} (1-\bs v^2)^{1/2}\;
f_t(\bs  v)
=
\f{t}{\tau} \,
\f{f_t(\bs  v)}{\gc(\bs v)},
\ee
where $\gamma(\bs v)=(1-\bs v^2)^{-1/2}$ is the Lorentz factor. We now define stationary distributions
\bse
\begin{eqnarray}
{f}_\infty({\bs  v})
=
\lim_{t\to \infty}f_t(\bs v),
\quad
\hat{f}_\infty({\bs  v})
=
\lim_{\tau\to\infty}
\hat{f}_\tau(\bs v).
\end{eqnarray}
Analogous to Eq.~\eqref{e:relation}, these equalities are to be understood in a distributional sense, i.e., 
\be
\int d^dv\; f_\infty (\bs v)\,g(\bs v)
&=&
\lim_{t\to \infty}
\int d^dv\; f_t(\bs v)\,g(\bs v),
\\
\int d^dv\; \hat{f}_\infty (\bs v)\,g(\bs v)
&=&
\lim_{\tau\to \infty}
\int d^dv\; \hat{f}_\tau (\bs v)\,g(\bs v)
\ee
\ese
for any sufficiently well-behaved, physically relevant test function $g(\bs v)$. If the dynamics is such that the limits $f_\infty$ and $ \hat{f}_\infty$ exist, then Eq.~\eqref{e:relation} implies that
\bse\label{e:stationary_relation}
\be
\hat{f}_\infty(\bs v)=\ga^{-1}\; f_\infty(\bs v)/\gc(\bs v),
\ee
where the constant 
\be
\ga= \int d^dv \;f_\infty(\bs v)/\gc(\bs v)
\ee
\ese
ensures normalization. Equation~\eqref{e:stationary_relation} states that, asymptotically, the $t$-averaged distribution  ${f}_\infty$ differs from  the $\tau$-averaged distribution  $\hat{f}_\infty$ by a factor proportional to the relativistic particle energy  $\eps=m\gc(\bs v)$, where $m$ is the rest mass of the particle.  We shall return to this point when discussing the associated entropy functionals further below. Before doing so, however, let us compare the time-averaged PDFs with \lq ensemble-averaged\rq\space one-particle velocity PDFs as recently measured in computer experiments~\cite{2007CuEtAl,2008MoGhBa,2006AlRoMo,2009DuHaHi}.

%\paragraph{Ensemble-averaged one-particle distributions.--}
{\em Ensemble-averaged one-particle distributions.--}
Molecular dynamics simulations by different groups~\cite{2007CuEtAl,2008MoGhBa,2006AlRoMo,2009DuHaHi} confirm that the stationary  one-particle velocity PDF of a $d$-dimensional dilute relativistic gas in equilibrium is  accurately described by the J\"uttner distribution~\cite{1911Ju,2008De}
\begin{equation}
\label{eq:juttner}
f_\J({\bs  v})
=Z_\J^{-1} m^{d}\gamma(\bs v)^{2+d}\exp[-\beta m\gamma(\bs v)], \qquad 
|\bs v|<1.
\end{equation}
Here, $T=(\kB \beta)^{-1}$ may be interpreted as a (rest) temperature, $\kB$  denotes the Boltzmann constant, and $Z_\J=Z_\J(d,m,\gb)$ the normalization constant. It is worthwhile to take a closer look at how exactly the measurements are performed in these simulations: 
\par
(i) Velocities are measured in the rest frame $\Gs_0$ of the boundary in the case of an enclosed system, or the center-of-mass frame in the case of periodic boundary conditions. 
\par
(ii) The velocities of all particles are measured  $t$-simultaneously in $\Gs_0$, where $t$ is the coordinate-time of~$\Gs_0$. 
\par
This procedure can be interpreted as constructing the one-particle velocity PDF via  \emph{ensemble-averaging}~\footnote{Simultaneous measurements can be easily made in computer simulations, but are are very difficult to perform in real experiments due to the finiteness of signal speeds in relativity; see discussion in~\cite{1967Ga,2009DuHaHi}.}. In the next part,  we would like to  compare the results of this method with those obtained by \emph{time-averaging} over a single-particle trajectory.

%\paragraph{Numerical simulations.--}
{\em Numerical simulations.--}
In order to understand how ensemble-averaged and time-averaged PDFs are related to each other,  we consider the fully relativistic (1+1)-dimensional two-component gas model studied in~\cite{2007CuEtAl}. In this model, the gas consists of classical, impenetrable point-particles ($N_1$ light particles of rest mass $m_1$, and $N_2$ heavy particles of rest mass $m_2>m_1$).  
Neighboring particles may exchange momentum and energy in elastic binary collisions, governed by the relativistic energy-momentum conservation laws
\be\notag
\eps(m_A,p_A)+\eps(m_B,p_B)&=&\eps(m_A,\tilde p_A)+\eps(m_B,\tilde p_B),\\
p_A+p_B&=&\tilde{p}_A+\tilde{p}_B,
\label{e:collision_kinematics}
\ee
where $A,B\in \{1,2\}$, $\eps(m,p)=(m^2+p^2)^{1/2}$, $p=m v\gc(v)$, and tilde-symbols indicate quantities after the collision. Interactions with the boundaries are elastic, i.e.,  $p\to -p$ in the lab frame $\Gs_0$,  defined as the rest frame of the boundaries. The dynamics of this system can be  exactly  integrated numerically, and the total energy $\HamEn_0=\sum_{i=1}^{N_1}\eps(m_1,p_i)+\sum_{j=1}^{N_2}\eps(m_2,p_j)$ is conserved in the lab frame $\Gs_0$. We distinguish four types of measurements.
\begin{itemize}
\item [(a)]
$t$-ensemble average: After a period of equilibration,  we simultaneously measure the velocities of all particles in $\Gs_0$ at a given instant of time $t$. This procedure is repeated for energetically equivalent random initial conditions mimicking a micro-canonical ensemble at energy $\HamEn_0$~\cite{2007CuEtAl}.
\item [(b)]
$t$-trajectory average: We choose a specific particle of either species and measure their velocities at several equidistant instants of time $t^{(1)},\ldots ,t^{(n)}$.
\item[(c)]
$\tau$-ensemble average: 
We compute the proper-time $\tau_i$  for each particle  during the simulation and measure their velocities at a fixed proper-time value $\tau_1=\ldots =\tau_{N_1+N_2}=\tau$. Again, this procedure is repeated for energetically equivalent random initial conditions.
\item[(d)]
$\tau$-trajectory average: 
We choose a specific particle of either species and measure their velocities at several equidistant  instants of proper-time $\tau^{(1)},\ldots ,\tau^{(n)}$.
\end{itemize}

\begin{figure}[t]
\centering
\includegraphics[width=7.5cm,angle=0]{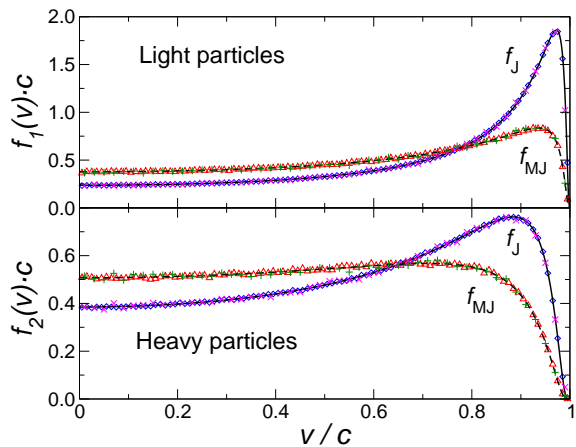}
\caption{
\label{fig:1} 
(Color online) Numerically measured one-particle velocity PDFs using  lab time and proper-time parameterizations, respectively. Symbols/methods: (a) $t$-ensemble average: blue $\diamond$,  (b) $t$-trajectory average: magenta $ \times$,  (c) $\tau$-ensemble average: red $\triangle$, and (d) $\tau$-trajectory average: green $+$.
The results are based on simulations with $N_1=5000$ light particles of mass $m_1$ and $N_2=5000$ heavy particles with mass $m_2=2m_1$. The solid curves correspond to J\"uttner functions~(\ref{eq:juttner}) with $\beta=0.709\, (m_1c^2)^{-1}$, but different particle masses, respectively.  Dashed lines show the corresponding modified J\"uttner distribution, Eq.~\eqref{e:ergodicity_tau}, with \emph{same} parameter $\gb$. As the distributions are symmetric with respect to the origin, only the positive velocity axis is shown.
}
\end{figure}

%\paragraph{Results.--}
{\em Results.--}
Figure~\ref{fig:1} depicts the equilibrium distributions computed from one-dimensional simulations as described above. In the case of the ensemble measurements (a) and (c)  we averaged over 50 different, energetically equivalent initial conditions. The single-particle time-averages (b) and (c) were determined by measuring velocities at $5\cdot10^5$ instants using time intervals  $\Delta t=\Delta \tau=4\cdot10^{-4}L/c$, where $L$ is the system's spatial extension. 
\par
Let us first compare the distribution functions obtained by the two $t$-averaging methods (a) and (b), respectively.  As evident from the (blue) diamonds and (magenta) cross-symbols in Fig.~\eqref{fig:1}, the two different procedures both yield a J\"uttner distribution with same parameter $\gb$ for either species, i.e.,
\bse\label{e:ergodicity}
\be\label{e:ergodicity_t}
f_\infty(v)=f_\J(v).
\ee
Similarly,  upon comparing the histograms obtained by (c) $\tau$-ensemble averaging,  see red triangles,  and (d) $\tau$-trajectory averaging, green plus-symbols,  we find that both methods give the same distribution. But this proper-time equilibrium PDF  differs from the J\"uttner function by a factor $1/\eps$, i.e.,
\be\label{e:ergodicity_tau}
\hat{f}_\infty(v)=\ga^{-1}  f_\J(v)/\gc(v)=:f_\MJ(v).
\ee
\ese
Thus, on the one hand, our simulations confirm the validity of Eq.~\eqref{e:stationary_relation} for the one-dimensional two-component gas model. One the other hand,  Eqs.~\eqref{e:ergodicity} provide two \lq soft\rq~ergodicity statements on the level of the one-particle velocity distributions (we adopt the term \lq soft\rq~rather than \lq weak\rq, which is already commonly used in a different context~\cite{1998KaHe}). Evidently, it is necessary to distinguish different time-parameters 
when discussing ergodicity in relativistic systems.
\par 
The \lq$\tau$-stationary\rq\space modified J\"uttner function~\eqref{e:ergodicity_tau} was derived earlier in Ref.~\cite{2007DuHa} from a  simple collision invariance criterion. Yet another  derivation, based on symmetry and entropy arguments,  was given in Ref.~\cite{2007DuTaHa_2}. However,  at that time it was not understood that the two distributions $f_\J$ and $f_\MJ$ refer to different time parameters,  respectively. In fact,  combining the above results with the arguments given in~\cite{2007DuTaHa_2} reveals an interesting relation  between time parameters and (relative) entropy in special relativity.

%%%%%%%%%%%%%%%%%%%%%%%%%%%%%%%%%%%%%%%
%%%%%%%%%%%%%%%%%%%%%%%%%%%%%%%%%%%%
%\paragraph*{Maximum (relative) entropy principle.--}
{\em Maximum (relative) entropy principle.--}
To establish a connection between $t$, $\tau$ and entropy, let us again generalize to the case of $d$-dimensional gas. We shall assume that gas consists of $N$ identical particles that  are enclosed by a vessel such that their total energy $\HamEn_0$ is conserved in the rest frame $\Gs_0$ of the vessel. For convenience, we rewrite the velocity distributions $f_\J$ and $f_\MJ$ in terms of the momentum coordinate $\bs p=m\bs v \gc(\bs v)$ as
\bse\label{e:momentum_distributions}
\be
\label{e:juttner-momentum}
\phi_\J(\bs p;\gb) &=& Z_\J^{-1} \exp(-\gb p^0),\\
\phi_\MJ(\bs p;\gb) &=& Z_\MJ^{-1} \phi_\J(\bs p;\gb)/ p^0,
\label{e:juttner-momentum_mod}
\ee
\ese
where  $p^0=(m^2+\bs p^2)^{1/2}$ is the relativistic one-particle energy, and $Z_\MJ=\ga Z_\J$. Written in this form, it becomes obvious that the distributions~\eqref{e:momentum_distributions} can be obtained by maximizing the relative entropy functional~\cite{2007DuTaHa_2}
\bse\label{e:entropy_principle}
\be\label{e:entropy_principle-a}
S[\phi|\rho_s]=
- \int\diff^dp\; \phi(\bs p) \ln\biggl[ \f{\phi(\bs p)}{\rho_s(\bs p)}\biggr],
\ee
under the constraints
\be
\label{e:juttner_constrain_1}
1&=&\int\diff^dp\; \phi(\bs p),\\
\f{\HamEn_s}{N}&=&\int\diff^dp\; \phi(\bs p)\,p^0,
\label{e:juttner_constrain_2}
\ee
\ese
{where $\HamEn_s/N$ is the specific energy mean value at constant coordinate-time $(s\equiv t)$ or proper-time  $(s\equiv \tau)$, respectively (cf. discussion below). } The function $\rho_s(\bs p)>0$ in Eq.~\eqref{e:entropy_principle-a} plays the role of a reference density~\cite{1976Ochs,1976Ochs_2,1951KuLe,1978We,1991We}  and ensures that the argument of the logarithm is dimensionless.  In 1911, J\"uttner~\cite{1911Ju} obtained the distribution $\phi_\J$ by postulating a constant reference density $\rho_t(\bs p)=\rho_0$, corresponding the translation-invariant Lebesgue measure in momentum space~\footnote{More precisely, J\"uttner~\cite{1911Ju} considered a Lebesgue reference measure on relativistic one-particle phase space $\{(\bs x,\bs p)\}$; however,  we can neglect the trivial spatial part of the distribution in our discussion.}. For comparison, the modified distribution $\phi_\MJ$ is obtained by fixing a reference density $\rho_\tau(\bs p) \propto 1/p^0$~\cite{2007DuTaHa_2}. It is well-known that 
$$
\diff \mu= \f{\diff^dp}{p^0}
$$
defines a unique (up to multiplicative factors) Lorentz-invariant integration measure in relativistic momentum space. Since $\phi_\MJ$ is related to the Lorentz-invariant time parameter $\tau$, we may conclude:  \emph{The symmetry properties of the reference density $\rho_s$, which appears in the relative entropy functional for the stationary distribution,  reflect the symmetry of the underlying time parameter.}
\par
{
It is worthwhile to briefly comment on the relation between the two different energy "state variables" $\HamEn_t$ and $\HamEn_\tau$  in Eqs.~\eqref{e:juttner_constrain_2}.
As evident from Eqs.~\eqref{e:stationary_relation} and also illustrated in Fig.~\ref{fig:1}, 
the inverse temperature parameter $\gb$ is the same for both the $t$-stationary distribution  $\phi_\J$ and the $\tau$-stationary distribution~$\phi_\MJ$. If we choose the maximum (relative) entropy principle as the starting point for deriving $\phi_\J$ and $\phi_\MJ$, the parameter $\gb$ enters as a Lagrangian multiplier for the energy constraint~\eqref{e:juttner_constrain_2}. Due to the elastic collision dynamics, the total initial energy~$\HamEn_0$ is conserved in the lab frame $\Gs_0$, if one measures temporal evolution in terms of the coordinate-time $t$. This suggests to identify $\HamEn_t=\HamEn_0$ in the case of the constant reference density~$\rho_t(\bs p)=\rho_0$, which yields the $t$-stationary  J\"uttner function $\phi_{\J}$. Upon inserting $\phi_\J$ into the energy constraint~\eqref{e:juttner_constrain_2}, one then finds that the parameter $\gb$ is uniquely determined by
\bse\label{e:relations}
\be\label{e:beta_condition}
\f{\HamEn_0}{N}&=&-\f{\p}{\p \gb} \ln Z_\J,\\
Z_\J&=& 2 m^d  \left(\f{2 \pi}{\gb m}\right)^{(d - 1)/2} K_{(d + 1)/2}(\gb m),  
\quad
\ee
where $d=1,2,3$ is the space dimension and $K_n(z)$ denotes the $n$th modified Bessel function of the second kind~\cite{AbSt72}.  Since, for any given initial energy value $\HamEn_0$, the parameter  $\gb$ is fixed by Eq.~\eqref{e:beta_condition}, it only remains to specify the corresponding "proper-time state variable"~$\HamEn_\tau$, which  is to be used in Eq.~\eqref{e:juttner_constrain_2} when considering a Lorentz-invariant reference density $\rho_\tau(\bs p) \propto 1/p^0$ in the relative entropy definition~\eqref{e:entropy_principle-a}. In general, the quantities $\HamEn_\tau$ and $\HamEn_t$ differ from each other, because they correspond to averages at $\tau=const.$ and $t=const.$, respectively. $\HamEn_\tau$~can be found by inserting the maximizer of $S[\phi|\gr_\tau]$, i.e., the
modified J\"uttner distribution~$\phi_\MJ(\bs p;\gb)$, into the energy constraint~\eqref{e:juttner_constrain_1}. This leads to\footnote{In $d=2$ space dimensions, Eq.~\eqref{e:beta_condition_2} can be rewritten as $\HamEn_\tau=N(m+1/\gb)$ which coincides with the classical non-relativistic equipartition theorem (unfortunately, this \lq coincidence\rq~holds only for $d=2$, but not for  $d=1,3$).}
\be\label{e:beta_condition_2}
\f{\HamEn_\tau}{N}&=&\f{Z_\J}{Z_\MJ},\\
Z_\MJ&=& 2 m^{d-1}  \left(\f{2 \pi}{\gb m}\right)^{(d - 1)/2} K_{(d - 1)/2}(\gb m).
\quad
\ee
\ese
Thus, by virtue of Eqs.~\eqref{e:relations}, each energy value $\HamEn_t=\HamEn_0$ corresponds uniquely to a  temperature value $\Temp=(\kB\gb)^{-1}$ and a "proper-time energy"~$\HamEn_\tau$.
Generally, when using a maximum (relative) entropy principle based on reference measures that refer to different time parameters, it is important to keep in mind that the "left-hand-side" of the energy constraint -- in our example, the parameter $\HamEn_s$ in Eq.~\eqref{e:juttner_constrain_2} -- must be specified in accordance with the underlying time parameter~$s$.
}

%\paragraph{Applications.--}
\pagebreak
{\em Outlook: Applications \& extensions.--}
{
At this point one may well ask whether or not different time-parameterizations and relativistic MD-type algorithms are relevant in practice. Generally, quasi-classical MD schemes can be useful for simulating the thermalization of relativistic gases and plasmas at sufficiently low densities. In the high-density regime more sophisticated methods are required that take into account quantum effects~\cite{2002BlIa, 2006HeGrRa_2,2000AaBoWe,2001AaBoWe,2002BlIa,2005Be} in a more detailed manner. For space dimensions $d>1$, classical particle-particle interactions cannot be formulated in a fully relativistic way anymore~\cite{1984MaMuSu} - unless one tried to keep track of interaction fields and retardation effects which is too expensive numerically. However, by choosing particle collision criteria carefully, e.g., by evaluating collisions in the two-body center-of-mass frame and using effective cross-sections as obtained from quantum theories~\cite{WeinbergQFT1}, one can achieve an acceptable degree of accuracy in low-to-moderate density simulations~\cite{2009DuHaHi}. 
}
\par
{
When applying MD methods to relativistic problems, coordinate-time parameterizations provide the more suitable framework for the causal ordering of collision events~\cite{1989SoStGr}. On the other hand, a proper-time parameterization is the more natural choice if one wishes to model decay processes on a classical particle level. This suggests to implement hybrid algorithms that combine deterministic MD schemes for the particle motion with randomized particle decay, annihilation, or creation. As a straightforward  extension of the above algorithm, one could account for such events by 
including chemical reactions that transform one or more particles into another (set of) particle(s). 
For example, in the simplest case, the proper-time intervals between successive decay events  can be sampled from an exponential distribution whose mean life-time is determined by quantum field theory.  Generally, the creation, reaction, and decay rules have to be chosen in agreement with the probabilitities and constraints (i.e., conservation laws) derived from the corresponding field theories~\cite{WeinbergQFT1}.  
}
\par
{
With regard to future investigations, numerical approaches appear to be particular promising when combined with novel GPU programming tools~\cite{2009PrEtAl,2009GeEtAl,2009JaKo,2009ThEtAl}. Compared with conventional CPU-based computations, GPU simulations techniques can significantly reduce computational costs (up to two orders of magnitude)  of parallelizable MD or MC algorithms. An example of the latter class are  relativistic MC simulations similar to those performed by Peano et al.~\cite{2009PeEtAl}, who generated the relativistic J\"uttner distribution~\eqref{eq:juttner} using randomized collisions. Their algorithm could be easily generalized to also include quantum phenomena such as, e.g., decay processes. In this context, it is  worth mentioning that the energy distribution of an unstable particle at the end of its life-time (assuming the particle lives long enough to allow for thermalization) will be described by  $\phi_\MJ$ rather than~$\phi_\J$.  
}

%\paragraph{Conclusions.--}
{\em Conclusions.--}
The above discussion shows that, depending on the application at hand, it may be useful to distinguish different types of stationarity in relativistic systems. Our numerical results for the one-dimensional  two-component gas illustrate that the J\"uttner distribution~\cite{1911Ju} $\phi_\J\propto\exp(-\gb p^0)$ is linked to coordinate-time $t$, whereas the modified distribution~ $\phi_\MJ\propto\exp(-\gb p^0)/p^0$ is linked to proper-time $\tau$. They further demonstrate that coordinate-time-averaging along a single-particle trajectory is equivalent  to $t$-simultaneous ensemble averaging over many particles. This may be interpreted as a soft form of $t$-ergodicity on the level of the one-particle velocity PDF in this model. An analogous statement holds for proper-time-parameterizations (soft $\tau$-ergodicity).
Moreover, the deterministic gas model provides a \lq microscopic\rq~illustration of conceptually similar results that were recently obtained within the theory of  relativistic Brownian motions~\cite{2009DuHaWe}. Similar to deterministic processes, relativistic Brownian motions can either be parameterized in terms of the coordinate-time $t$ or their proper-time~$\tau$.  The resulting stationary momentum distributions (if existing) are then connected by a relation equivalent to Eq.~\eqref{e:stationary_relation}. This result is, perhaps, more difficult to understand (or accept) when considering stochastic processes based on postulated random driving processes. The deterministic model system considered here helps to clarify the microscopic origin of this relativistic effect. Last but not least, our analysis implies a close connection between  entropy and time parameters, which may be summarized as follows: A  Lorentz-invariant time parameter corresponds to a Lorentz-invariant reference density (probability measure) in the entropy functional for the associated stationary distribution.

%\paragraph{Acknowledgements.--}
\acknowledgments
This research was supported by the Ministerio de Ciencia e Innovaci\'on of Spain, project number FIS2008-02873 (D.C.), and the Junta de Andalucia (D.C).

\bibliographystyle{eplbib}
%\bibliography{BrownianMotion,BrownianMotors,BlackHoles,Cosmology,CosmicRays,QuantumBM,Debbasch,DiverseBooks,Einstein,Entropy,Ergodicity,FinanceBooks,FT,Hakim,Hanggi,LorentzDirac,Numerics,PathIntegrals,PhotonDiffusion,RelKin,RelStatMech,RelMany,RelConstraint,RelFPE,RelStochQuant,RelTD,RBM,RBMapplied,RBMappliedHE,RBMappliedAstro,RBMradiation,SpecRelativity,StochProc,Telegraph,VanKampen,TD,Journals,Proceedings,MathBooks,ActiveBM,Anomalous,Schimansky,Ebeling,Theses,Selfenergy,GPU_examples,NonEqQCD}

%%%%%%%%%%%%%%%%%%%%%%%%%%%%%%%%%%%%%%%%%%%%%%%%%%
%%%%%%%%%%%%%%%%%%%%%%%%%%%%%%%%%%%%%%%%%%%%%%%%%%
%%%%%%%%%%%%%%%%%%%%%%%%%%%%%%%%%%%%%%%%%%%%%%%%%%

\end{document}